\documentclass[12pt]{article}

\usepackage{epsfig}
\usepackage{amsmath}
\usepackage{amssymb}
\usepackage{graphicx}
\usepackage{cite}
\usepackage{enumerate}
  \newcommand{\abar}{\bar{\alpha}_s}

 \newcommand{\beq}{\begin{eqnarray}}
  \newcommand{\eeq}{\end{eqnarray}}

\def\independenT#1#2{\mathrel{\setbox0\hbox{$#1#2$}%
 \copy0\kern-\wd0\mkern4mu\box0}} 
\def\permil{\%\raise.10ex\hbox{$_{\scriptstyle 0}$}}
\def\bea{\begin{eqnarray}}
\def\eea{\end{eqnarray}}
\def\bsp{\begin{split}}
\def\esp{\end{split}}

\def\be{\begin{equation}}
\def\ee{\end{equation}}
\def\bea{\begin{eqnarray}}
\def\eea{\end{eqnarray}}
\def\bsp{\begin{split}}
\def\esp{\end{split}}
\begin{document}
%\titlepage
\titlepage
\begin{flushright}
IFJPAN-IV-2011-13 \\
\end{flushright}
\vspace*{1in}
\begin{center}
{\Large \bf Nonlinear equation for coherent gluon emission.}\\
\vspace*{0.5cm}
Krzysztof Kutak, Krzysztof Golec-Biernat, Stanislaw Jadach, Maciej Skrzypek \\
\vspace*{0.5cm}
{\it Instytut Fizyki J\c{a}drowej im. H. Niewodnicza\'nskiego,\\
Radzikowskiego 152, 31-342 Krak\'ow, Poland\\}
\end{center}
\vspace*{1cm}
\centerline{(\today)}

\vskip1cm
\begin{abstract}
Motivated by the regime of QCD explored nowadays at LHC, where both the
total energy of collision and momenta transfers are high, we investigate evolution equations
of high energy factorization. In order to study such effects like parton saturation in final
states one is inevitably led to investigate how to combine physics of the BK and CCFM
evolution equations. In this paper we obtain a new exclusive form of the BK equation
which suggests a possible form of the nonlinear extension of the CCFM equation.
\end{abstract}

\section{Introduction}
The Large Hadron Collider (LHC) is already operational and Quantum Chromodynamics (QCD) is the basic theory which 
is used to set up the initial conditions for the collisions at the LHC as well as  to calculate hadronic observables. 
The application of perturbative QCD relies on the so called factorization theorems which allow to decompose a given process 
into a long distance part, called parton density, and a short distance part, called matrix element. Here we will focus on 
high energy factorization \cite{Catani:1990eg,Gribov:1984tu}. The evolution equations of high energy factorization sum up 
logarithms of energy accompanied by a strong coupling constant, i.e. terms proportional to $\alpha_s^n \ln^m s/s_0$, which 
applies when the total energy of a scattering process is much bigger than any other hard scale involved in a process.

Until now, in principle, the BFKL \cite{Kuraev:1977fs,Balitsky:1978ic,Mueller:1994jq}, BK \cite{Balitsky:1995ub,Kovchegov:1999yj,Kovchegov:1999ua} 
and CCFM \cite{Ciafaloni:1987ur,Catani:1989sg,Catani:1989yc} evolution equations were used on equal footing since the energy ranges 
did not allow to discriminate between these frameworks. However, there 
were indications already at HERA \cite{Stasto:2000er} for the need to account for nonlinear effects in gluon density. These observation are 
supported by recent results obtained 
in \cite{Albacete:2010pg,Dumitru:2010iy}. On top of this, the results from \cite{Deak:2010gk} point at the need to use the framework which incorporates hardness of the collision into BFKL like description.
With the LHC one entered into a region of phase space where both the energy and
momentum transfers are high. Therefore, one should provide a framework where both
dense systems and hard processes can be studied. This might be achieved with a relatively
simple nonlinear extension of the CCFM equation where one could take into account both
the gluon production and recombination in the description of final states.

In this paper we obtain such an equation,  see eq. (\ref{eq:final1}). In order to arrive at this equation in the first 
crucial step we perform resummation of virtual and unresolved real contributions in the BK equation in a similar fashion as in 
 \cite{Kwiecinski:1996td}. 
As a result, we arrive at a new form of the BK see eq.(\ref{eq:bkexclusive}) equation where both the linear and nonlinear parts are folded with a Regge form 
factor in which singularities of the linear part have been resumed.
In this new form, the singularity in the unresolved real contribution of the $s$-channel real gluon is canceled by the virtual 
contribution. This is the minimal condition in order to eventually perform a Monte Carlo simulation based 
on the BK equation.

However, the BK equation concerns inclusive observables and does not allow for applications to the description 
of exclusive final states. Thus, we need the CCFM equation which is applicable to the description of the exclusive processes.  
We propose the nonlinear extension of the CCFM equation being motivated by the resumed form of the BK equation in which we replace 
the Regge form factor by the non-Sudakov form factor and introduce angular ordering (coherence). In addition we supplement the BK kernel with a large $x$ part.
For a different approach to the extension of the CCFM equation to allow for gluon saturation we refer the reader to 
\cite{Kutak:2008ed,Avsar:2009pf,Avsar:2010ia}. For an approach in which inter-jet observables are resummedd by nonlinear evolution equation which has an analogous structure as in the BK equation we refer the reader to 
\cite{Banfi:2002hw,Marchesini:2003nh,Weigert:2003mm}
and paper where the exact mapping was found \cite{Hatta:2008st}.   

The paper is organized as follows. In section 2 we introduce the BK evolution equation
for the dipole amplitude in the momentum space and perform resummation of unresolved
real emissions and virtual emissions arriving at a new representation of the BK equation.
In section 3 we present the main result of this paper which is a new evolution equation
using the resummed BK equation i.e. - the CCFM equation extended by a nonlinear term.

\section{Exclusive form of the Balitsky-Kovchegov equation}
\begin{figure}[!t]
\centerline{\epsfig{file=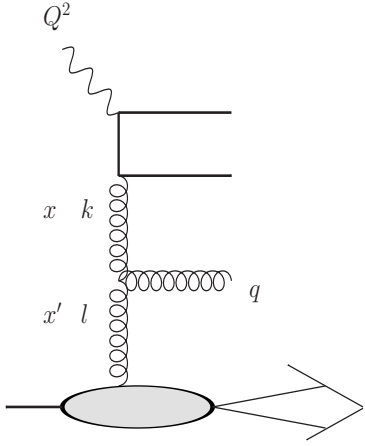,height=6cm,width=5cm}} 
\caption{{\it Plot explaining meaning of variables in BK and CCFM evolution equations.}}
\label{fig:Kinematics}
\end{figure}

At the leading order in $\ln 1/x$ the Balitsky-Kovchegov equation for the dipole amplitude in the momentum space is \cite{Kovchegov:1999ua}: 
\begin{align}
 \frac{\partial\Phi(x,k^2)}{\partial \ln 1/x}=\overline\alpha_s\int_0^{\infty}\frac{dl^2}{l^2}
\bigg[\frac{l^2\Phi(x,l^2)- k^2\Phi(x,k^2)}{|k^2-l^2|}+ \frac{
k^2\Phi(x,k)}{\sqrt{(4l^4+k^4)}}\bigg]-\overline\alpha_s\Phi^2(x,k)\,.
\label{eq:faneq0}
\end{align}
the linear term can be linked to the process of creation of gluons while the nonlinar term
can be linked to fusion of gluons and therefore introduces gluon saturation effects. In order
to find an exclusive form of the BK equation and define a link to the CCFM equation, in
the first step we rewrite it as an integral equation following the KMS framework\cite{Kwiecinski:1997ee}
\begin{align}
\label{eq:fan1}
\Phi(x,k^2)&= \Phi_0(x,k^2)\\\nonumber
&+\overline\alpha_s\int_{x}^1 \frac{dz}{z}\int_0^{\infty}\frac{dl^2}{l^2}
\bigg[\frac{l^2\Phi(x/z,l^2)- k^2\Phi(x/z,k^2)}{|k^2-l^2|}+ \frac{
k^2\Phi(x/z,k)}{\sqrt{(4l^4+k^4)}}\bigg]\\\nonumber
&-\overline\alpha_s\int_{x}^1 \frac{dz}{z}\Phi^2(x/z,k)
\nonumber
\end{align}
where the  lengths of transverse vectors lying in transversal plane to the collision axis are $k\equiv|{\bold k}|$, $l\equiv|{\bold l}|$ (${\bold k}$ is a 
vector sum of transversal momenta of emitted gluons during evolution), $z=x/x'$(see Fig. (\ref{fig:Kinematics}), $\overline\alpha_s=N_c\alpha_s/\pi$. The impact parameter dependence $b$ of the 
dipole amplitude in momentum space is assumed to be trivial i.e. in a form of a theta function, $\theta(R-b)$ with $R$ defining 
the target radius. Therefore we suppress it but it is understood implicitly.
The unintegrated gluon density obeying the high energy factorization theorem \cite{Catani:1990eg} is obtained from
 \cite{Braun:2000wr,Kutak:2003bd,Kutak:2004ym}: 
\be
{\cal F}_{BK}(x,k^2)=\frac{N_c}{\alpha_s \pi^2}k^2\nabla_k^2 \Phi(x,k^2)
\ee
where the angle independent Laplace operator is given by 
$\nabla_k^2=4\frac{\partial}{\partial k^2}k^2\frac{\partial}{\partial k^2}$.\\
In order to arrive at an exclusive form of the BK equation first we  
reintroduce the angular dependence to the linear part of eq. (\ref{eq:fan1}). We obtain
\begin{align}
\label{eq:bkversion1}
\Phi(x,k^2)&=\Phi^{0}(x,k^2)\\\nonumber
&+\overline\alpha_s\int_x^1\frac{dz}{z}\int\frac{d^2{\bf q}}{\pi q^2}\left[\Phi(x/z,|{\bold k}+{\bold q}|^2)-
\theta(k^2-q^2)\Phi(x/z,k)\right]\\\nonumber
&-\overline\alpha_s\int_x^1\frac{dz}{z}\Phi^2(x/z,k)\nonumber
\end{align}
where ${\bf q}={\bf l}-{\bf k}$.
%%%%%%%%%%%%%%%%%%%%%%%%%%%%%%%%%%%%%%%%%%%%%%%%%%%%%%%%%%%%%%%%%%%%%%%%%%
Introducing the resolution scale $\mu$ and decomposing the linear part of eq. (\ref{eq:bkversion1}) into resolved real emission part with $q^2>\mu^2$
and the unresolved part with $q^2<\mu^2$, we obtain
\begin{align}
\label{eq:bkversion2}
\Phi(x,k^2)&=\Phi^{0}(x,k^2)\\\nonumber
&+\overline\alpha_s\int_x^1\frac{dz}{z}\int\frac{d^2{\bf q}}{\pi q^2}\Phi(x/z,|{\bold k}+{\bold q}|^2)\theta(q^2-\mu^2)\\\nonumber
&+\overline\alpha_s\int_x^1\frac{dz}{z}\int\frac{d^2{\bf q}}{\pi q^2}\big[\Phi(x/z,|{\bold k}+{\bold q}|^2)\theta(\mu^2- q^2)-
\theta(k^2-q^2)\Phi(x/z,k)\big]\\\nonumber
&-\overline\alpha_s\int_x^1\frac{dz}{z}\Phi^2(x/z,k)\,.
\nonumber
\end{align}
A convenient way of performing resummation or exponentiation of virtual and unresolved real emissions is provided by the Mellin transform defined as
\be
\overline\Phi(\omega,k^2)=\int_0^1 dx x^{\omega-1} \Phi(x,k^2)
\ee
while the inverse transform reads
\be
\Phi(x,k^2)=\int_{c-i\infty}^{c+i\infty} d\omega\, x^{-\omega} \overline\Phi(\omega,k^2)\,.
\label{eq:invmellin}
\ee
Performing the Mellin transform and using in the unresolved part $|{\bold k}+{\bold q}|^2\approx{\bold k}^2 $ since $q^2<\mu^2$ we obtain
\begin{align}
\label{eq:eqtransform}
\overline \Phi(\omega,k^2)&=\overline \Phi^{0}(\omega,k^2)\\\nonumber
&+\frac{\overline\alpha_s}{\omega}\int\frac{d^2{\bold q}}{ q^2}[\overline \Phi(\omega,|{\bf k} +{\bf q}|^2)\theta(q^2-\mu^2)]
+\frac{\overline\alpha_s}{\omega}\int\frac{d^2{\bold q}}{ q^2}{\overline \Phi}(\omega,k^2)[\theta(\mu^2-q^2)-\theta(k^2-q^2)]\\\nonumber
&-\frac{\overline\alpha_s}{\omega}\int_0^1 dy y^{\omega-1}\Phi^2(y,k^2)
\end{align}
where in the nonlinear term we changed the variables $x/z\rightarrow y$ and we integrated over $x$ what gives $1/\omega$ in front of the nonlinear part.
After combining the unresolved real and virtual parts we obtain
\begin{align}
\overline \Phi(\omega,k^2)&=\overline \Phi^{0}(\omega,k^2)\\\nonumber
&+\frac{\overline\alpha_s}{\omega}\int\frac{d^2{\bf q}}{\pi q^2}\overline \Phi(\omega,|{\bf k} +{\bf q}|^2)\theta(q^2-\mu^2)-\frac{\overline\alpha_s}{\omega}\overline \Phi(\omega,k^2)\ln\frac{k^2}{\mu^2}\\\nonumber
&-\frac{\overline{\alpha}_s}{\omega}\int_0^1 dy y^{\omega-1}\Phi^2(y,k^2)\,.
\nonumber
\end{align}
This can be simplified to
\begin{align}
\overline \Phi(\omega,k^2)&=\hat \Phi^{0}(\omega,k^2)\\\nonumber 
&+\frac{\overline\alpha_s}{\overline\omega+\omega}\int\frac{d^2{\bf q}}{\pi q^2}\overline \Phi(\omega,|{\bf k} +{\bf q}|^2)]\theta(q^2-\mu^2)-\frac{\overline{\alpha}_s}{\omega+\overline\omega}\int_0^1 dy y^{\omega-1}\Phi^2(y,k^2)
\end{align}
where 
\be
\hat \Phi^{0}(\omega,k^2)=\frac{\omega\,\overline\Phi^{0}(\omega,k^2)}{\omega+\overline\omega}\,,~~~~~~~~
\overline\omega=\overline\alpha_s\ln\frac{k^2}{\mu^2}.
\ee
The inverse transform (\ref{eq:invmellin}) can be computed as follows
\begin{align}
\Phi(x,k^2)&=\frac{1}{2\pi i}\int_{c-i\infty}^{c+i\infty} d\omega\,x^{-\omega}\hat\Phi^{0}(\omega,k^2)\\\nonumber 
&+\frac{1}{2\pi i}\int_{c-i\infty}^{c+i\infty} d\omega\,x^{-\omega}\frac{\overline\alpha_s}{\omega+\overline\omega}\int\frac{d^2{\bf q}}{\pi q^2}\int_0^1 dy\,y^{\omega-1}\Phi(y,|{\bf k} +{\bf q}|^2)]\\\nonumber
&-\frac{1}{2\pi i}\int_{c-i\infty}^{c+i\infty} d\omega
x^{-\omega}\frac{\overline{\alpha}_s}{\omega+\overline\omega}\int_0^1 dy\, y^{\omega-1}\Phi^2(y,k^2)\,.
\end{align}
Calculating the residue, changing variables $y\rightarrow x/z$ and using $\Delta_R(z,k,\mu)\equiv\exp\left(-\overline\alpha_s\ln\frac{1}{z}\ln\frac{k^2}{\mu^2}\right)$, which is called Regge 
form factor, we obtain:
\begin{eqnarray}
\Phi(x,k^2)\!\!\!&=&\!\!\!\tilde \Phi^0(x,k^2)+\overline\alpha_s\int\frac{d^2{\bf q}}{\pi q^2}\int_x^1\frac{d\,z}{z}\Delta_R(z,k,\mu)\,\Phi(\frac{x}{z},|{\bf k} +{\bf q}|^2)\,\theta(q^2-\mu^2)
\\\nonumber
&-&\!\!\!\int_x^1\frac{dz}{z}\Delta_R(z,k,\mu^2)\,\Phi^2(\frac{x}{z},k^2)
\end{eqnarray}
where we introduced
\be
\tilde\Phi^0(x,k^2)\equiv \frac{1}{2\pi i}\int_{c-i\infty}^{c+i\infty} d\omega\,x^{-\omega}\hat\Phi^{0}(\omega,k^2)\,.
\ee
Finally we obtain
\begin{align}
\Phi(x,k^2)&=\tilde \Phi^0(x,k^2)\\\nonumber
&+\overline\alpha_s\int_x^1\frac{d\,z}{z}\Delta_R(z,k,\mu^2)\Big[\int\frac{d^2{\bf q}}{\pi q^2}\,\theta(q^2-\mu^2)\,\Phi(\frac{x}{z},|{\bf k} +{\bf q}|^2)-\Phi^2(\frac{x}{z},k^2)\Big]\nonumber
\end{align}
or equivalently
\begin{eqnarray}
\label{eq:bkexclusive}
\Phi(x,k^2)\!\!\!&=&\!\!\!\tilde \Phi^0(x,k^2)\\
&+&\!\!\!\overline\alpha_s\int_x^1d\,z\int\frac{d^2{\bf q}}{\pi q^2}\,
\theta(q^2-\mu^2)\frac{\Delta_R(z,k,\mu)}{z}\left[\Phi(\frac{x}{z},|{\bf k} +{\bf q}|^2)-q^2\delta(q^2-k^2)\,\Phi^2(\frac{x}{z},q^2)\right].
\nonumber
\end{eqnarray}
Eq. (\ref{eq:bkexclusive}) is a new form of the BK equation in which the resummed terms in a form of 
Regge form factor are the same for the linear and nonlinear part. 
This form will serve as a guiding equation to generalize the CCFM equation to include nonlinear effects which allow for recombination of partons.
It will also be useful as a starting point in an attempt to solve nonlinear extension of the CCFM  equation using iterative and Monte Carlo 
methods \cite{GolecBiernat:2007xv}.

\section{Towards nonlinear extension of the CCFM equation}
Motivated by the suggestive form of the BK equation where the Regge form factor appears in the nonlinear part, in this section we propose 
an extension of the CCFM equation to account for the nonlinearity, what could be viewed as an extension of the BK equation to the large $x$ domain. This extension is of course to some extend arbitrary since at present we 
do not have diagrammatic picture of the triple pomeron vertex \cite{Bartels:1994jj,Bartels:2007dm} which would account for coherence. 
Our guiding principle is eq. (\ref{eq:bkexclusive}) in which the linear part resembles the CCFM equation. 
The expectation is that at the low $x$ limit the solution of nonlinear extension of the CCFM equation should approach the solution of the BK equation. Also they should give similar predictions for inclusive observables.\\
\subsection{CCFM evolution equation}
The CCFM equation sums up gluonic emissions with the condition of strong ordering in emission angle which allow for smooth interpolation between BFKL limit when $z\rightarrow\,0$ 
and the DGLAP limit (in gluonic channel) when $z\rightarrow 1$. The BFKL or the BK are generally applied for the calculation of elastic
scattering and total cross-sections while due to the dependence on the hard scale (which is expressed as the maximal 
angle) the CCFM equation allows for studies of exclusive observables.
It is the following
%\begin{multline}
\begin{align}
\label{eq:ccfmorg}
\mathcal{A}(x, k^2, p) &=  \mathcal{A}_0(x,k^2,p)\\\nonumber
&+\abar \int_x^1 dz
\int \frac{d^2\bar{\bf{q}}}{\pi \bar{q}^2} \,\theta (p - z\bar{q})
\Delta_s(p,z\bar{q})
\left ( \frac{\Delta_{ns}(z,k, q)}{z} + \frac{1}{1-z} \right )
\mathcal{A}\left(\frac{x}{z}, k^{'2}, \bar{q}\right).
\\\nonumber
\nonumber
\end{align}
The gluon density obtained from the CCFM equation, usually denoted $\mathcal{A}(x,k^2,p)$, on a level of linear equation  has interpretation of the gluon density 
describing parton with longitudinal momentum fraction 
$x$, and transverse momentum (squared) $k^2$ which is probed by a hard system at scale $p$.\\  
The momentum vector associated with $i$-th emitted gluon is
\be
q_i=\alpha_i\,p_P+\beta_i\,p_e+q_{t\,i}
\ee
from which the rapidity and angle of emitted gluon with respect to incoming parent proton (beam direction) can be obtained as
\be
\eta_i=\frac{1}{2}\ln(\xi_i)\equiv\frac{1}{2}\ln\left(\frac{\beta_i}{\alpha_i}\right)=\ln\left(\frac{|{\bf
q}_{i}|}{\sqrt{s}\,\alpha_i}\right),\,\,\,\,\,\tan\frac{\theta_i}{2}=\frac{|{\bf q}_i|}{\sqrt s\,\alpha_i}\,.
\ee
The variable $p$ in (\ref{eq:ccfmorg}) is defined via $\bar{\xi} = p^2/(x^2s)$ where $\frac{1}{2}\ln(\bar{\xi})$ is a maximal rapidity which is determined by the kinematics of hard scattering,
$\sqrt{s}$ is the total energy of the collision. For example, for the electron proton scattering $s=(p_P+p_e)^2$ and $k' = |\pmb{k} + (1-z)\bar{\pmb{q}}|$.
Using the variables $\xi$ the angular ordering can be conveniently expressed as: $\bar{\xi}\!>\!\xi_i\!>\!\xi_{i-1}\!>\!...\!>\!\xi_1\!>\!\xi_0$, where $\xi_0\equiv\mu$ with $\mu$ being infrared cut off.
The momentum ${\bf\bar{q}}$ is the transverse rescaled momentum of the real gluon, and is related 
to ${\bf q}$ by $\bar{{\bf q}} = {\bf q}/(1-z)$ and $\bar q\equiv|{\bar{\bf q}}|$.\\ 
The Sudakov form factor which screens the $1-z$ singularity is given by
\be
\Delta_s(p,z\bar q)=\exp\left(-\bar\alpha_s\int_{(z\bar{q})^2}^{p^2}\frac{dq^2}{q^2}\int_0^1\frac{dz}{1-z}\right).
\ee
The form factor $\Delta_{ns}$ screens the $1/z$ singularity, in a similar form as the Regge form factor 
but also accounts for angular ordering:
\be
\Delta_{ns}(z,k,q)=\exp\left(-\overline\alpha_s\int_
z^1\frac{dz'}{z'}\int_{z^{\prime 2}q^2}^{k^2}\frac{dq^{\prime 2}}{q^{\prime 2}}\right)=\exp\left(-\alpha_s\ln\frac{1}{z}\ln\frac{k^2}{z q^2}\right).
\label{eq:nonsudakov}
\ee
where for the lowest value of $z q^2$ we use a cut off $\mu$. In the literature \cite{Kwiecinski:1995pu}
different forms of the non-Sudakov form factor were considered, where the differences came from the approximations taken, the one we are using 
follows results obtained in \cite{Catani:1989sg}. 
For studies of effects of various forms of non-Sudakov form factors we refer the reader to \cite{Avsar:2010ia}. For our purposes it is enough to
 mention that the nonlinearity which suppresses contributions from gluons with low transversal momenta makes the dependence on the non-Sudakov's
  form subleading. This is because the non-Sudakov form factor affects mainly the low $k$ part of the gluon distribution.

\subsection{Nonlinear extension of the CCFM equation}
As it has already been stated the motivation to extend the CCFM to account for nonlinearity is to be able to study the impact of saturation of partons on exclusive observables. There are indications \cite{Kutak:2008ed,Adloff:1996dy} that such effects might be significant in for instance 
production of charged particles at HERA or in forward production of di-jets \cite{Albacete:2010pg}.\\
Below we propose the extension of the CCFM  equation to include nonlinearity. In order to avoid confusion we introduce new notation for the extended
 by nonlinearity the
CCFM equation. We also want to recall that the nonlinear extension of the CCFM changes the interpretation
of the quantity for which the equation is written. It is not longer high energy factorizable gluon density ${\mathcal A}$ but should be interpreted as the
dipole amplitude in momentum space $\Phi$, denoted from now on by ${\mathcal E}$,  which
depends additionally on a hard scale $p$. Below we propose the new equation, the peculiar structure of the nonlinear term is motivated by the following 
requirements:
\begin{itemize}
\item the second argument of the $\mathcal{E}$ should be $k^2$ as motivated by the analogy to BK
\item the third argument should reflect locally the angular ordering
\end{itemize}
\begin{align}
\label{eq:final1}
\mathcal{E}(x, k^2, p) &=  \mathcal{E}_0(x,k^2,p)\\\nonumber
&+\abar \int_x^1 dz
\int \frac{d^2\bar{\bf{q}}}{\pi \bar{q}^2} \,\theta (p - z\bar{q})
\Delta_s(p,z\bar{q})
\left ( \frac{\Delta_{ns}(z,k, q)}{z} + \frac{1}{1-z} \right )\Bigg[
\mathcal{E}\left(\frac{x}{z}, k^{'2}, \bar{q}\right)\\\nonumber
&-\bar{q}^2\delta(\bar{q}^2-k^2)\,\mathcal{E}^2(\frac{x}{z},\bar{q}^2,\bar{q})\Bigg].\nonumber
\end{align}
Similarly as in case of the BK equation in order to obtain high energy factorizable unintegrated gluon density one applies: 
\be
{\cal A}_{non-linear}(x,k^2,p)=\frac{N_c}{\alpha_s \pi^2} k^2\,\nabla_k^2\, \mathcal{E}(x,k^2,p)\,.
\ee
The nonlinear term  in (\ref{eq:final1}),  apart from allowing for recombination 
of gluons might be understood as a way to introduce the decoherence into the emission pattern of gluons. 
This is because the gluon density is build up due to coherent gluon emission and since the nonlinear term 
comes with the negative sign it slows down the growth of gluon density and therefore it introduces the decoherence.\\
We expect the nonlinear term to be
of main importance at low $x$ similarly as in case of the BK equation. In this limit it  will be of special interest to check whether in this formulation of the nonlinear 
extension of the CCFM equation one obtains an effect of saturation of the saturation
scale as observed in \cite{Avsar:2010ia}. In the mentioned paper it has beens stated that extension of the CCFM equation which mimics saturation by special boundary conditions 
generates saturation scale which saturates itself due to constraints on the phase space coming from maximal allowed scale by the kinematics. 
This effect is of great importance since it has a consequences for example for imposing a bound on amount of production of entropy from saturated part of 
gluon density as observed in \cite{Kutak:2011rb}. 
Another interesting limit is the large $x$ region of the phase space where the nonlinear effects are believed to be subleading and can be safely neglected. However,
since the CCFM extends beyond BFKL asymptotics, with the presented formulation one can study onset of saturation in a function of $x$.

%%%%%%%%%%%%%%%%%%%%%%%%%%%%%%%%%%%%%%%%%%%%%%%%%%%%%%%%%%%%%%%%%%%%%%%%% 
\section{Conclusions and outlook}
In this paper we have studied the high energy factorizable evolution equations. We obtained a new form of the BK equation where the 
unresolved real and virtual contributions in the linear part of the kernel are resummed. On top of this, both the linear part and
nonlinear pieces became folded with the Regge form factor. The obtained representation
of the BK equation might likely be useful in future Monte Carlo solution. However, in
order for it to be useful for phenomenological studies of exclusive final states one has to
take also into account angular ordering of gluon emissions. This we do by extending the
CCFM equation by nonlinear term following the suggestive form of the BK equation in the
exclusive representation. In a future, the obtained new equation will be investigated with
a focus on possibility to solve it using Monte Carlo methods.

\section*{Acknowledgments}
We would like to thank M. Deak, H. Jung, F. Hautmann and W. P\l aczek for usefull discussions.
This research has been supported by the grant Homing Plus/2010-2/6 of Fundacja na rzecz Nauki Polskiej and the grant of 
Polish National Science Center no.~DEC-2011/01/B/ST2/03915.

\end{document}